\documentclass{aastex63}
\usepackage{gensymb}
\usepackage{amssymb}

\received{April 30, 2021}
\revised{}
\accepted{}
\submitjournal{Planetary Science Journal: THAI Focus Issue}

\shorttitle{ExoCAM}
\shortauthors{Wolf et al.}

\graphicspath{{./}{figures/}}

\begin{document}

\title{ExoCAM: A 3D Climate Model for Exoplanet Atmospheres}

\correspondingauthor{Eric Wolf}
\email{eric.wolf@colorado.edu}

\author[0000-0002-0786-7307]{Eric T. Wolf}
\affiliation{University of Colorado, Boulder \\
Laboratory for Atmospheric and Space Physics, \\
 Department of Atmospheric and Oceanic Sciences, \\
 Boulder, CO} 
\affiliation{NASA GSFC Sellers Exoplanet Environments Collaboration, Greenbelt, MD}
\affiliation{NASA NExSS Virtual Planetary Laboratory, Seattle, WA}

\author[0000-0002-5893-2471]{Ravi Kopparapu}
\affiliation{NASA Goddard Space Flight Center \\
8800 Greenbelt Road \\
Greenbelt, MD 20771, USA}
\affiliation{NASA GSFC Sellers Exoplanet Environments Collaboration}

\author[0000-0003-4346-2611]{Jacob Haqq-Misra}
\affiliation{Blue Marble Space Institute of Science,\\
Seattle, WA, USA}

\author[0000-0002-5967-9631]{Thomas J. Fauchez}
\affiliation{NASA Goddard Space Flight Center \\
8800 Greenbelt Road \\
Greenbelt, MD 20771, USA}
\affiliation{Goddard Earth Sciences Technology and Research (GESTAR), Universities Space Research Association (USRA), Columbia, MD 7178, USA}
\affiliation{NASA GSFC Sellers Exoplanet Environments Collaboration}

\begin{abstract}

The TRAPPIST-1 Habitable Atmosphere Intercomparison (THAI) project was initiated to compare 3D climate models that are commonly used for predicting theoretical climates of habitable zone extrasolar planets.  One of the core models studied as part of THAI is ExoCAM, an independently curated exoplanet branch of the National Center for Atmospheric Research (NCAR) Community Earth System Model (CESM) version 1.2.1. ExoCAM has been used for studying atmospheres of terrestrial extrasolar planets around a variety of stars.  To accompany the THAI project and provide a primary reference, here we describe ExoCAM and what makes it unique from standard configurations of CESM.  Furthermore, we also conduct a series of intramodel sensitivity tests of relevant moist physical tuning parameters while using the THAI protocol as our starting point.  A common criticism of 3D climate models used for exoplanet modeling is that cloud and convection routines often contain free parameters that are tuned to the modern Earth, and thus may be a source of uncertainty in evaluating exoplanet climates.  Here, we explore sensitivities to numerous configuration and parameter selections, including a recently updated radiation scheme, a different cloud and convection physics package, different cloud and precipitation tuning parameters, and a different sea ice albedo.  Improvements to our radiation scheme and the modification of cloud particle sizes have the largest effect on global mean temperatures, with variations up to $\sim$10~K, highlighting the requirement for accurate radiative transfer and the importance of cloud microphysics for simulating exoplanetary climates.   However for the vast majority of sensitivity tests, climate differences are small.  For all cases studied, intramodel differences do not bias general conclusions regarding climate states and habitability.

\end{abstract}

\vspace{5mm}
\section{Introduction} \label{sec:intro}
\section{ExoCAM Description} \label{sec:description}

As introduced in section~\ref{sec:intro}, ExoCAM is a modeling package designed to be used in tandem with the National Center for Atmospheric Research Community Earth System Model version 1.2.1 to facilitate flexible modeling of extrasolar planetary atmospheres.  In recent years, interest in 3D exoplanet atmosphere modeling has exploded, but the road to proficiency for new users entering into the field is often steep and arduous.  Furthermore, the real-time and computing-time required for full-cycle science (project genesis, model development, simulations, analysis, publication) with 3D climate models means that current 3D exoclimate modeling groups are often already near or at their carrying capacity and cannot contribute meaningfully to all new spontaneous collaborations that may emerge.  The goal of ExoCAM is to lower the barrier for entry into the field for new users, by providing an accessible platform to begin climate modeling of planets and exoplanets while leveraging the NCAR CESM framework.   ExoCAM provides accessibility for idealized planet configurations, which are most typical of exoplanet studies, with relatively straight-forward handling of modifications of the basic geophysical, orbital, and atmospheric parameters of the modeled planet, and a flexible radiative transfer code.  Note that ExoCAM and ExoRT are continually updated on GitHub when new features are developed or bugs are fixed.  We encourage users to stay up-to-date with the code, and to contact the authors for guidance regarding model features, model capabilities, and appropriate use strategies.

\subsection{General Features} \label{subsec:generalfeatures}

To begin using ExoCAM, it is recommended that a new user first gain a general understanding and proficiency of core CESM code for which there is extensive documentation provided by NCAR\footnote{https://www.cesm.ucar.edu/models/cesm1.2/cesm/doc/usersguide/ug.pdf}, along with a robust user forum\footnote{https://bb.cgd.ucar.edu/cesm/} for troubleshooting common problems.  CESM combines component models of the Community Atmosphere Model (CAM)\footnote{https://www.cesm.ucar.edu/models/ccsm4.0/cam/docs/description/cam4$\_$desc.pdf},\footnote{https://www.cesm.ucar.edu/models/cesm1.2/cam/docs/description/cam5$\_$desc.pdf}, the Community Land Model (CLM)\footnote{https://www.cesm.ucar.edu/models/cesm1.2/clm/CLM4$\_$Tech$\_$Note.pdf}, the Community Ice Code (CICE)\footnote{https://www.cesm.ucar.edu/models/cesm1.2/cice/ice$\_$usrdoc.pdf}, options for either a slab ocean model\footnote{https://www.cesm.ucar.edu/models/cesm1.2/data8/doc/ug.pdf} or the Parallel Ocean Program (POP)\footnote{https://www.cesm.ucar.edu/models/cesm1.2/pop2/doc/sci/POPRefManual.pdf} dynamic ocean model, and a River Transport Model (RTM), which are all coupled together\footnote{https://www.cesm.ucar.edu/models/cesm1.2/cpl7/doc/ug.pdf}.  Users guides and technical descriptions are included via the footnotes above.  CESM can be configured in multiple ways, combining, some or all of the functionalities of the above model components depending on the user's goals for their simulations.  After basic proficiency is gained with CESM, ExoCAM can be downloaded and applied as an upgrade, following the instructions provided on the GitHub pages. 

Presently, ExoCAM supports several standard configurations, horizontal resolutions, and vertical layerings out-of-the-box.  ExoCAM supported configurations include a q-flux aquaplanet slab ocean \citep{bitz:2012}, a completely land covered planet, and a mixed land and slab ocean world.  Surface boundary data sets are provided for an aquaplanet with zero heat transport, a completely land covered planet, and for an Earth-continental configuration with prescribed q-flux ocean heat transports that mimic the present day. ExoCAM also comes with Interactive Data Language (IDL) scripts with basic functionality for changing surface boundary conditions files, initial condition files, and vertical layerings.  

ExoCAM can be run with either the CAM4 cloud and convection physics \citep{rasch&kristjansson:1998, hack:1994, zhang&mcfarlane:1995}, or with the CAM5 cloud and convention physics \citep{morrison&gettelman:2008, park&bretherton:2009, park:2014}.  To date, all published ExoCAM simulations have used the CAM4 cloud and convection physics, due to its simpler form and numerically more robust performance for strongly forced atmospheres \citep{bin:2018}.  The CAM5 cloud physics provides a more sophisticated two-moment scheme that can capture the interaction between clouds and aerosols; however, it requires assumptions to be made regarding the background aerosol fields.  Continuing studies of cloud and aerosol interactions on exoplanets and solar system objects should be a promising area of future research using ExoCAM.  Both CAM4 and CAM5 convection routines use mass-flux approaches \citep{ooyama1971theory} to parameterize the convection. In such a scheme, the depth of the convection is constrained by the distance the air, rising in convective updrafts, penetrates above its level of neutral buoyancy.  Note that the \citet{zhang&mcfarlane:1995} deep convection scheme has been modified to use the more robust numerical solver of \citet{brent:1973} (Courtesy C.A. Shields), which improves numerical stability for warm moist climate states.  For further reading, detailed and specific descriptions of the CAM4 and CAM5 cloud and convection physics are provided in the associated technical manuals \citep{neale:2010a, neale:2010b} and references within.   

The primary dynamical core used with ExoCAM is a finite volume (FV) scheme \citep{lin&rood:1996}.  However, a supported configuration is also offered to use the spectral element (SE) dynamical core on a cubed sphere grid \citep{lauritzen:2014}.  The FV dynamical core has been modified to improve numerical stability in more strongly forced atmospheres by incrementally applying physics tendencies for temperature and wind speed evenly throughout the dynamical substeps, rather than only at the beginning of the dynamics step \citep{bardeen:2017}.  Several studies \citep{lebonnois:2012, lauritzen:2014} found that NCAR's FV dynamical core does not properly conserve angular momentum for slow rotating planets and inadequately captures upper-level superrotation for Venus and Titan atmospheres \citep{larson:2014}.  \citet{lauritzen:2014} recommends the use of the SE dynamical core for simulating worlds such as Venus and Titan.  However, \citet{kopparapu:2017} conducted tests comparing the resultant climate states for slowly rotating aquaplanets around M-dwarf stars using both the FV and SE dynamical cores, and found that the surface climates and runaway greenhouse tipping points were unaffected by the choice of dynamical core between FV and SE options. Still, exploring the effects of the SE dynamical core for capturing upper level supperrotation on tidally locked planets remains a promising area for future research using ExoCAM.  

ExoCAM simulations are typically run at the fairly coarse horizontal resolution of 4\degree$\times$5\degree.  However, other researchers have successfully run ExoCAM at resolutions of 1.9\degree$\times$2.5\degree, 0.5\degree$\times$0.5\degree \citep{wei:2020}, and 0.47\degree$\times$0.63\degree \citep{komacek:2019} respectively. Note that the choice of horizontal resolution should be commensurate with the relevant length scales of the large scale circulation features, which typically depend on the size of planet and its rotation rate, combined with other factors \citep{haqq-misra:2018}.  For terrestrial exoplanet studies at Earth-like rotation rates and slower, 4\degree$\times$5\degree resolution is suitable.  Indeed, 4\degree$\times$5\degree resolution and similar are commonly used.  However, for faster rotating planets one may need to increase the horizontal resolution to maintain integrity of the atmospheric dynamics (e.g. \citet{komacek:2019}).

Likewise, ExoCAM can be easily run with variety of vertical layers.  A 40 layer vertical grid, spanning 3 orders of magnitude in pressure, is the default selection provided with ExoCAM.  Nevertheless, a secondary grid with 51 layers spanning 5 orders of magnitude is also provided (e.g. \citet{suissa:2020a}).  Note, however, numerical instability issues tend to become more prevalent as one extends the model top to lower pressures, often requiring reduced timestep and substep lengths.  Note, other researchers have successfully run ExoCAM with the CESM default 26 layer grid, and also a 71 layer grid \citep{wei:2020}.  IDL scripts are provided to perform basic manipulations of the vertical layerings in ExoCAM.

Finally, ExoCAM provides basic controls for setting the geophysical and atmospheric properties of the modeled planet, along with several options for controlling modes of operation.  While CESM-centric options and settings continue to be set by the native namelist files, ExoCAM-centric options are set at build-time within the source file exoplanet$\_$mod.F90.  Note, that most CESM-centric options set in namelist files do not require recompiling the model; however, parameters set in exoplanet$\_$mod.F90 require recompiling the model when changes are made. The file exoplanet$\_$mod.F90 contains settings for planet radius, surface gravity, rotation rate and period, incident stellar flux, stellar spectrum, orbital eccentricity, obliquity, and partial pressure of atmospheric gases.  The specific heat of dry air and molecular weight of the atmosphere are set automatically at build-time based on the assigned partial pressures of atmospheric gases and thus do not need to be specified by the user.  Note, initial condition files are provided for a variety of surface pressures which must be selected to match the total pressure set in exoplanet$\_$mod.F90.  Additionally, there are a series of boolean options to enable certain functionalities, such as setting the planet to synchronous rotation, conducting a parallel clear-sky radiative calculation to compute cloud forcings, outputting of spectrally resolved radiative fluxes, and enabling prognostic gravity waves.

\subsection{Radiative Transfer} \label{subsec:radiation}
At the heart of ExoCAM is its radiative transfer package, ExoRT.  Any flexible planetary model must be able to handle radiative transfer for a wide variety of atmospheric compositions.  Native Earth-centric radiation schemes tend to perform poorly when pushed outside the bounds of Earth-similar atmospheres.  The success of ExoCAM is thus critically dependent on providing a versatile radiative transfer package, and thus the largest share of historical development effort over the years has gone into improving ExoRT.  ExoRT is provided via GitHub as a standalone package which is linked to ExoCAM at build-time.  ExoRT uses a correlated-k distribution approach, and the two-stream approximation of \citet{toon:1989}.  The code lineage traces back to \citet{urata&toon:2013a,urata&toon:2013b} and \citet{colaprete:2003} with extensive modifications and reorganization occurring over time.  There are now several radiative transfer options available for use with ExoCAM.  While the most recent update is now recommended for all use cases moving forward, we describe all available versions below. 

The oldest radiation version is named \textit{n28archean} in the ExoRT repository, and is originally described in \citep{wolf&toon:2013}. This radiative transfer version features 28 spectral intervals across the full spectrum ranging from 10 to 50000~cm$^{-1}$ (0.2 to 1000~$\mu$m).  Nominally, the longwave calculation is computed over 17 bins between 10 and 4000 cm$^{-1}$ (2.5 to 1000~$\mu$m), while the shortwave calculation is computed over 23 bins between 980 and 50000~cm$^{-1}$ (0.2 to 10.2~$\mu$m).  Note that in \textit{n28archean}, and all other configurations described below, the shortwave bandpass is extended far into the thermal wavelengths in order to accommodate TRAPPIST-1-like spectral energy distributions out-of-the-box.  However, the exact bandpasses for the longwave and shortwave streams are modifiable by the user within file exo$\_$init$\_$ref.F90 at compile-time, and can be adjusted considering the stellar and planetary temperatures to squeeze out additional computational efficiency.  This is true for all ExoRT configurations.  Molecular absorption is included from H$_2$O, CO$_2$, and CH$_4$.  Correlated-k distributions were produced using the Atmospheric Environment Research Inc. line-by-line radiative transfer model (LBLRTM, \citet{clough:2005}) using the HITRAN2004 spectroscopic database \citep{rothman:2005}.  For all gases we assumed Voigt line wings cut at 25~cm$^{-1}$.  Water vapor and CO$_2$ continuum absorption are included using MT$\_$CKD version~2.5, while assuming that CO$_2$ self-broadened continuum is 1.3 times the foreign broadening continuum \citep{kasting:1984, halevy:2009}.  Overlapping gas absorption is handled using an amount weighted scheme, with no more than two overlapping in each band \citep{mlawer:1997}. Collision induced absorption (CIA) is included for N$_2$-N$_2$, N$_2$-H$_2$, H$_2$-H$_2$ pairs.   This version was originally designed for studying the climate of Archean Earth, for CO$_2$ amounts up to several tenths of a bar and up to several thousand ppm of CH$_4$ along with Earth-like water vapor amounts.  Results were initially compared to line-by-line calculations (see Figure 1 in \citet{wolf&toon:2013}) for Archean-like atmospheres. Note that this version of the radiative transfer is used for the official THAI simulations discussed in \citet{turbet:2021, sergeev:2021} and \citet{fauchez:2021b}.

In 2016 an intercomparison project was initiated by \citet{yang_j:2016} to study the performance of various radiative transfer codes used for simulating warm, moist terrestrial planet atmospheres, up to 360~K.  This study found that \textit{n28archean} (referred to as "CAM4$\_$Wolf" in \citet{yang_j:2016}) performed well in the longwave; however, in the limit of warm moist atmospheres and M-dwarf stellar spectra, it significantly overestimated the strength of near-infrared absorption of the downwelling shortwave radiation stream, resulting in too much radiation being absorbed by the atmosphere and not enough reaching the surface.  Motivated by this study, ExoRT was updated to its second iteration, first described in \citet{kopparapu:2017} and referred to in the repository as \textit{n42h2o}.  This version increased the number of spectral intervals to 42 across the full spectrum ranging from 10 to 50,000~cm$^{-1}$ (0.2 to 1000~$\mu$m). Nominally, the longwave calculation is computed over 19 bins between 10 and 4000~cm$^{-1}$ (2.5 to 1000~$\mu$m), while the shortwave calculation is computed over 35 bins between 820 and 50000~cm$^{-1}$ (0.2 to 12.2~$\mu$m).  Spectral resolution was preferentially increased in the near-infrared region in order to address deficiencies discovered in \citet{yang_j:2016}).  This version includes only H$_2$O gas absorption using the HITRAN2012 \citep{rothman:2013} spectroscopic database along with the BPS water vapor continuum \citep{paynter&ramaswamy:2011}.  Correlated-k distributions for H$_2$O were produced using HELIOS-K, a GPU-accelerated k-distribution sorting code \citep{grimm&heng:2015}, which is freely available on GitHub\footnote{https://github.com/exoclime/HELIOS-K}.  Line wings assume a Voigt shape and a 25~cm$^{-1}$ cut-off. This radiative transfer scheme improved the model performance in the near-infrared around M-dwarf stars and served the purpose of facilitating studies of water vapor and water cloud interactions on tidally locked planets around M-dwarf stars.  CIA is included for N$_2$-N$_2$, N$_2$-H$_2$, H$_2$-H$_2$ pairs.  This scheme was compared favorably against LBLRTM for warm moist N$_2$, H$_2$O atmospheres (see Figure 1 in \citet{kopparapu:2017}).  Note also there is an analogous 68 spectral interval version of this radiation scheme, \textit{n68h2o}, which was used in \citet{wolf:2019} to yield better output spectral resolution for use in phase curve studies.  This version has a full spectrum range from 0 to 42087.00~cm$^{-1}$ (0.238 to infinity~$\mu$m) with longwave and shortwave  calculations computed over the entire spectral range.

In the fall of 2020, ExoRT received a major overhaul, which improves model performance and greatly improves the flexibility of the model to be able to incorporate a wider variety of overlapping gases.  This most recent version is called \textit{n68equiv}.  All users of ExoCAM are now recommended to use \textit{n68equiv}, superseding both \textit{n42h2o} and \textit{n28archean}.  This new version uses 68 spectral intervals across the full spectrum ranging from 0 to 42087.00~cm$^{-1}$ (0.238 to infinity~$\mu$m).  Nominally, the longwave calculation is computed over 37 bins between infinity and 4030~cm$^{-1}$ (2.48 to infinity~$\mu$m), while the shortwave calculation is computed over 53 bins between 875 and 42087~cm$^{-1}$ (0.238 to 11.43~$\mu$m).  Importantly, \textit{n68equiv} now uses the equivalent extinction method to handle overlapping gases \citep{amundsen:2017}.  In the equivalent extinction method, the primary gas absorber in each spectral interval is treated with an 8 Gauss point k-distribution, while all additional species that have absorption in the band are treated as grey absorbers using their band averaged absorption value.  The major species in each band is dynamically selected at each model grid cell and each model time-step by first comparing the band averaged optical depths of each species before the full radiative calculation is done.  The equivalent extinction method for gas overlap offers a significant advantage over other overlap schemes because it permits many overlapping species to be present in each spectral interval without requiring the prior computation of mixed gas k-coefficients tables, which become unwieldy if considering more than two gas species per interval. Presently, \textit{n68equiv} includes gas absorption from H$_2$O, CO$_2$, and CH$_4$, with k-distributions produced by HELIOS-K \citep{grimm&heng:2015} using the HITRAN2016 spectroscopic database \citep{gordon:2017}.  Voigt line shapes with a 25~cm$^{-1}$ line cut-off were used for H$_2$O and for CH$_4$.  For H$_2$O the plinth, or base-term \citep{paynter&ramaswamy:2011}, was removed to prevent double counting of absorption when combined with continuum absorption.  Self and foreign water vapor continuum coefficients are included using MT$\_$CKD version~3.2.  However, instead of using band-averaged values, we found that improved performance in the near-infrared region is achieved by fitting the water vapor self and foreign continuum coefficients to an 8-point k-distribution, and included under the assumption of perfect correlation when water vapor is present in a grid cell.  For CO$_2$ we assume the sub-Lorenztian line-shape of \citet{perrin&hartmann:1989} with a 500~cm$^{-1}$ line cut-off.  In addition to N$_2$, H$_2$ CIA pairs, \textit{n68equiv} includes CIA for CO$_2$-CO$_2$ \citep{wordsworth:2010} and for CO$_2$-H$_2$ and CO$_2$-CH$_4$ from \citet{turbet:2020}.   {Note that there is also an analogous 84 spectral interval version, \textit{n84equiv}, which is identical in description to \textit{n68equiv}, except 16 additional spectral intervals have been added shortward of 0.238~$\mu$m.  Testing indicates that for stars hotter than about 6500~K, non-negligible stellar flux is pushed shortward of $\sim$0.2~$\mu$m and was being effectively lost by the \textit{n68equiv}.  If using  \textit{n84equiv}, users are encouraged to examine and tailor the specific shortwave bandpass limits in exo$\_$init$\_$ref.F90 to balance the need for encompassing far-ultraviolet wavelength ranges with computational cost. 

Some features are shared by all available versions of ExoRT and are described here.  All versions treat liquid and ice cloud droplets as Mie scattering particles, with refractive indices taken from \citep{segelstein:1981} and \citet{warren&brandt:2008}.  The radiative effect of cloud overlap is treated using the Monte Carlo Independent Column Approximation while assuming maximum-random overlap \citep{pincus:2003, barker:2008}.  Rayleigh scattering is included with the parameterization of \citet{vardavas&carver:1984}, including the effects of N$_2$, CO$_2$, H$_2$O, and H$_2$.  The longwave and shortwave streams share a common spectral grid and k-coefficient tables.  The extent of the longwave and shortwave spectral integration streams can be readily adjusted to balance completeness versus computational speed, depending upon the stellar type assumed and planetary temperatures involved (see file exo$\_$init$\_$ref.F90). All k-coefficient sets currently use log$_{10}$(P$_1$/P$_2$)=0.1 spacing.  The \textit{n28archean} scheme has a pressure range from 100~bars to 0.01~mb, and a temperature range from 80~K to 520~K.  The \textit{n68equiv} and \textit{n42h2o} versions have a pressure range from 10~bars to 0.01~mb, and a temperature range from 100~K to 500~K,    All versions use the Gauss point binning originated from RRTMG \citep{mlawer:1997} that is weighted towards 1 and thus favors capturing the peak of the absorption lines in each band.  A variety of input stellar spectrum are provided for each radiative transfer version, ranging from TRAPPIST-1-like stars (2600~K) up to F-type stars (10000~K).  Additional software is provided, written in IDL for manipulating input and output data for ExoRT. 

Past papers, as noted, illustrate the performance of \textit{n42h2o} \citep{kopparapu:2017} and \textit{n28archean} \citep{wolf&toon:2013, yang_j:2016}.  Here, in Figures \ref{h2ort_fig} and \ref{co2rt_fig}, we illustrate the basic radiative transfer performance of \textit{n68equiv} against line-by-line calculations for standard atmospheres.  Recall, \textit{n68equiv} is the recommended version for all use cases; however, \textit{n28equiv} can still be used for modern Earth and Archean Earth cases while offering faster computation than \textit{n68equiv}, and \textit{n42h2o} remains valid, albeit while lacking greenhouse gases besides H$_2$O.  Figure \ref{h2ort_fig} duplicates the radiative transfer calculations of warm moist atmospheres originally conducted in in \citet{yang_j:2016}, including comparisons against LBLRTM \citep{clough:2005} and SMART \citep{meadows&crisp:1996, goldblatt:2013}, considering both the outgoing longwave radiation (OLR) at the top-of-atmosphere (TOA), and the downwelling stellar flux at the surface under irradiation from Solar and M-dwarf (3400~K blackbody) spectra respectively.  The x-axis tracks with the mean surface temperature.  These atmospheres are assumed to contain 1~bar N$_2$, 376~ppm of CO$_2$ and are assumed to be fully saturated with water vapor.  {Water vapor adds pressure to the atmosphere, thus by 360~K the total surface pressure is $\sim$1.6~bars.  For longwave fluxes, \textit{n68equiv} performs closely to LBLRTM, while both diverge from SMART for higher temperature H$_2$O-rich atmospheres.  This is likely because \textit{n68equiv} and LBLRTM use the same H$_2$O continuum treatment (MT$\_$CKD) while SMART uses a $\chi$-factor approach \citep{goldblatt:2013}.  The shortwave performance of \textit{n68equiv} is dramatically improved compared to \textit{n28archean} \citep{yang_j:2016}, but differences are still evident for hotter atmospheres, likely due to the complexity of the near-infrared water vapor spectrum, which is difficult to fully resolve at coarse spectral resolutions suitable for 3D climate model operations.

Figure \ref{co2rt_fig} shows comparisons of the outgoing longwave radiative flux between \textit{n68equiv}, SMART, and SOCRATES (a high-resolution band-model,  \citet{edwards&slingo:1996}) for a 2~bar pure-CO$_2$ dry atmosphere at Mars gravity with a 250~K surface temperature. SMART calculations shown were originally published in \citet{kopparapu:2013}. Both \textit{n68equiv} and SOCRATES yield OLR values of $\sim$93.3~Wm$^{-2}$, while SMART yields an OLR of $\sim$88.5~Wm$^{-2}$.  Here, \textit{n68equiv} notably lacks the spectral resolution to capture the fine detail of CO$_2$ absorption between 800 and 1200~cm$^{-1}$.  Still, the primary differences between \textit{n68equiv} and SOCRATES compared to SMART are most noticeable in the wing regions of the primary 667~cm$^{-1}$ CO$_2$ absorption band.  This may be attributed to differences in the line wing treatments between \citet{perrin&hartmann:1989} used in \textit{n68equiv} and SOCRATES, compared to stronger line-wing formulation used in SMART \citep{meadows&crisp:1996}.  See also, Figure 1 in \citet{halevy:2009} for a comparison of relevant CO$_2$ line wing treatments.  Note, several other studies have computed identical CO$_2$ radiation tests.  Identical column tests by \citet{wordsworth:2010} with a correlated-k band model yielded 88.2~Wm$^{-2}$, in line with SMART. Later \citet{wordsworth:2017} conducted line-by-line calculations which yielded $\sim$92~Wm$^{-2}$.  SOCRATES calculations featured in \citep{guzewich:2021} yielded 96~Wm$^{-2}$; however those calculations used an out-of-date CO$_2$-CO$_2$ CIA file which is missing absorption between 250 and 500~cm$^{-1}$.  Note, our goal here is not to anoint any one model to be better than the other, but rather to give the reader an adequate benchmarking of performance versus other popular models. Note that the \textit{n28archean} version of ExoRT, which has been used in prior ExoCAM studies of CO$_2$ dominated atmospheres \citep{wolf:2018a, wolf:2018b} and in the standard THAI simulations \citep{sergeev:2021, turbet:2021}} yields an OLR of 81.3~Wm$^{-2}$, an overestimation of the greenhouse effect for CO$_2$ dominated atmospheres. Future studies of CO$_2$ dominated atmospheres with ExoCAM should use \textit{n68equiv}.

\begin{figure}
  \centering
  \includegraphics[angle=0,width=12.5cm]{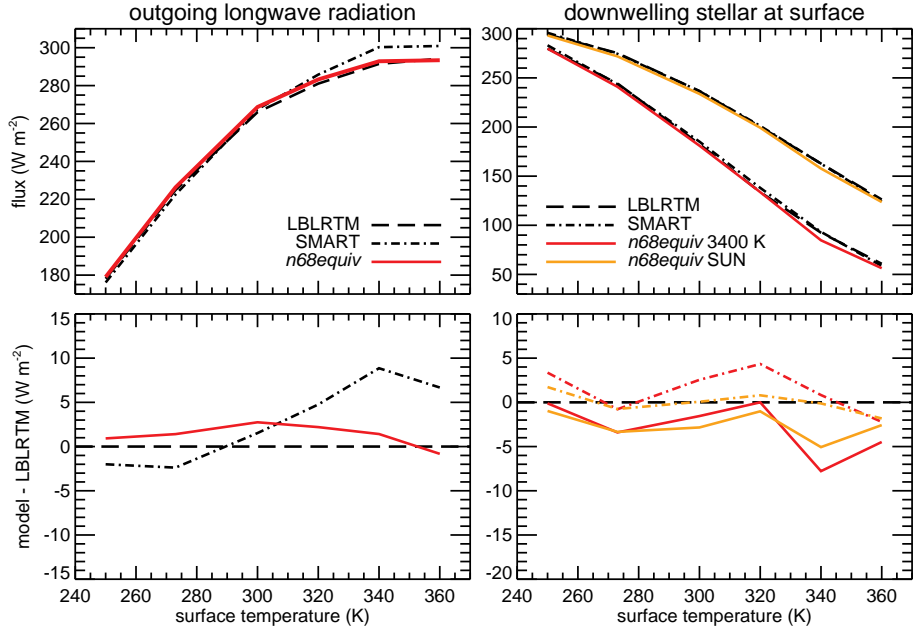}
  \caption{Here we show the outgoing longwave radiation and the downwelling stellar flux at the surface, compared between line-by-line calculations from SMART and LBLRTM, and our most recently updated radiative transfer version \textit{n68equiv}. The top row shows the flux results, while the bottom row shows the difference between \textit{n68equiv} and each line-by-line model, respectively.  We follow the procedures of \citet{yang_j:2016} and examine atmospheres that have surface temperatures according to the x-axis, with 1~bar N$_2$, 376~ppm CO$_2$, and are saturated with water vapor.  Our \textit{n68equiv} version compares favorably against line-by-line results, and marks a significant improvement over previous versions of ExoRT.}
  \label{h2ort_fig}
\end{figure}

\begin{figure}
  \centering
  \includegraphics[angle=0,width=12.5cm]{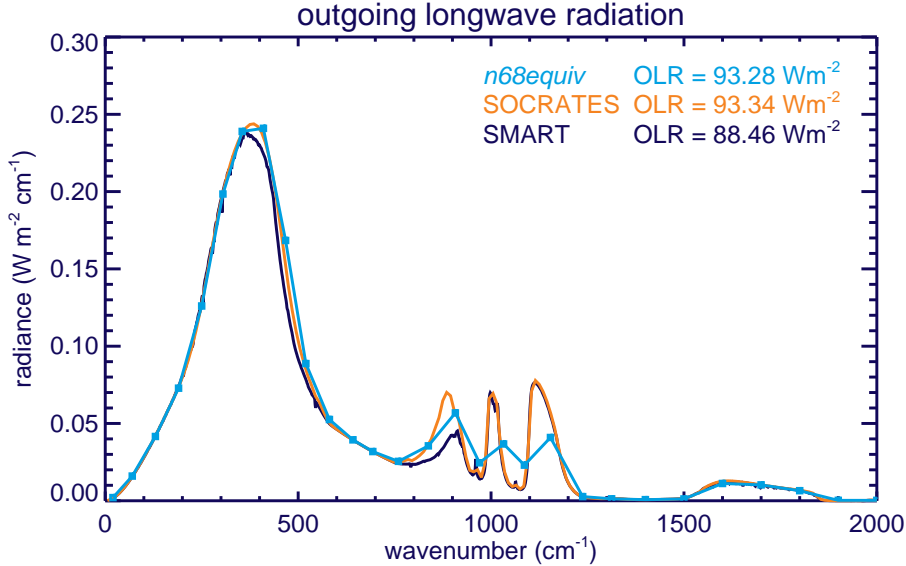}
  \caption{Here we show the spectral outgoing longwave radiation compared between line-by-line calculations with SMART, high resolution band model calculations with SOCRATES, and \textit{n68equiv}.  We assumed a 2~bar CO$_2$ atmosphere with a 250~K surface temperature, and no other absorbing species. SOCRATES and\ textit{n68equiv} are similar while SMART calculations, originally appearing in \citet{kopparapu:2013}, feature strong absorption.  Differences in total absorption strength are likely due to the line wing treatments \citep{perrin&hartmann:1989, meadows&crisp:1996}  }
  \label{co2rt_fig}
\end{figure}

\subsection{Additional Features} \label{subsec:additionalfeatures}
There are several additional functionalities that can be used as part of ExoCAM. We have implemented a circumbinary planet model based on the dynamical calculations of \citep{georgakarakos&eggl:2015} and featured in \citep{wolf:2020}.  Circumbinary systems are those where a planet orbits two stars on a long period orbit, while the two host stars orbit each other on a shorter period.  The orbital motions of the binary stars drive time-dependent changes in the stellar flux received by the planet.  Our circumbinary model explicitly resolves these changes including time-dependent, spectral, zenith angle, and binary eclipsing effects.  The circumbinary model presently is only offered with the \textit{n28archean} radiative transfer scheme, which in this case has been fully embedded in ExoCAM rather than linked.  The integration between the circumbinary orbital computer and ExoRT contains additional complications making the radiative transfer schemes not immediately swappable without additional redesign across all versions.  While there are no immediate plans to port the circumbinary orbital computer to other radiation versions, users interested in using the circumbinary model with the \textit{n68equiv} radiation scheme are encouraged to contact the lead author for further collaboration.

Source code is also provided that links ExoCAM and ExoRT with the Community Aerosol and Radiation Model for Atmospheres (CARMA) model \citep{turco:1979, toon:1988, bardeen:2008}.  CARMA is a flexible cloud and aerosol model that resides on the publicly supported trunk of CESM and can be flexibly configured to treat any manner of cloud and aerosol problem, including photochemical hazes \citep{wolf&toon:2010, larson:2014}, Martian ice clouds and dust \citep{hartwick:2019}, historical volcanic eruptions \citep{english:2013}, geoengineering \citep{english:2012}, nuclear winter \citep{mills:2008}, extinction level asteroid impacts \citep{bardeen:2017}, forest fires \citep{yu:2019} and a wide variety of other problems.  At the time of the finalizing of this paper, we offer a photochemical haze CARMA configuration \citep{wolf&toon:2010} linkable to our 28 spectral interval radiation (see \textit{n28archean.haze} in ExoRT); however, developing a more formal ExoCAM-CARMA integration with \textit{n68equiv} for multiple cloud and aerosol types is an active area of work in the coming years.

There are additional model components that exist within CESM that can be feasibly connected to ExoCAM, but are not yet offered formally as part of ExoCAM.  For instance, \citep{chen:2018, chen:2019, chen:2021} has used the ExoCAM orbital and geophysical control modules along with the native chemistry (Model for Ozone and Related Chemical Tracers, MOZART, \citet{kinnison:2007}) and upper atmosphere (Whole Atmosphere Community Climate Mode, WACCM, \citet{marsh:2007}) packages to explore the effects of M-dwarf stellar spectra and stellar activity, including energetic proton fluxes, on the atmospheres of tidally locked Earth-like exoplanets.  Furthermore, it is trivial to initialize ExoCAM with the POP2 dynamic ocean model instead of the slab ocean model, using the NCAR provided Earth topographic and bathymetric boundary conditions.  However, the user is warned that while it is simple to engage POP2 with Earth boundary conditions, it is non-trivial to change the ocean boundary condition data sets for POP2 and alternative sets are not currently provided.  Still, for the motivated user instructions for changing topographic and bathymetric boundary conditions are provided by NCAR as part of their Paleo Tool Kit\footnote{http://www.cesm.ucar.edu/models/paleo/faq/}.

\subsection{Computational Performance}
ExoCAM successfully compiles and runs on a wide variety of national and university machines.  Computational performance depends on many factors, including the horizontal resolution, the number of vertical layers, the number of advected constituents, the choice of radiative transfer scheme, the engagement of different model components, and the setting of various model timestepping and substepping iterations within the physics, dynamics, and radiation routines respectively. A typical out-of-the-box ExoCAM simulation is configured with the following setup: atmosphere, slab ocean, sea ice, and/or land model components enabled, a horizontal resolution of 4\degree$\times$5\degree with 40 vertical layers, 3 advected constituents (water vapor, ice water cloud, and liquid water cloud), a primary model timestep (namelist variable \textit{dtime}) of 30 model minutes, a dynamical substepping frequency (namelist variable \textit{nsplit}) of 32 per primary timestep, dynamical remapping and tracer advection frequencies (namelist variables \textit{nspltvrm} and \textit{nspltrac}) of 4 per primary timestep, with the radiative transfer called once every 2 model timesteps (exoplanet$\_$mod.F90 setting \textit{exo$\_$rad$\_$step}), and using the \textit{n68equiv} radiation scheme. Operated in such a mode, run on NASA's Discover supercomputer using Intel Xeon Haswell chips, one Earth year can be simulated in $\sim$180 processor-hours.  Slab ocean simulations take $\sim$50 model years to sufficiently complete, thus resulting in a total computational cost of about $\sim$9,000 core hours for a basic simulation to be run to equilibrium.  

Of course, each user's experience will vary depending on alterations to the above described configuration.  Naturally, horizontal and vertical spatial resolutions have a large impact on the computational performance.  The radiative transfer spectral resolution and the number of advected constituents tend to be expensive run-time components.  The recommended \textit{n68equiv} radiation scheme is notably expensive due to its relatively large number of spectral intervals. Running the model with CARMA or MOZART enabled, which entails numerous additional advected species and calculations, can slow the model down considerably.  Furthermore, the inclusion of deep atmospheres, probing near climatic tipping points, or configuring with the POP2 dynamic ocean model can require increasingly long model time frames to reach convergence, regardless of computational efficiency.

There are different methods that can be employed to squeeze out gains in computational performance that may be applicable depending on one's particular science requirements.  These include using fewer vertical levels, configuring with coarser horizontal resolutions, reducing the frequency of the radiative transfer calculations, compile-time tightening up the specific shortwave and longwave bandpasses in the radiative transfer (see module exo$\_$init$\_$ref.F90) to suite your specific star and planet temperature combination, reducing the number of dynamical substeps, and adjusting the primary model timestep. As a first approach, reducing the radiative transfer timestep can yield significant gains in performance since it is generally the most expensive component.  While our nominal radiative transfer timestep is 60 model minutes,  radiative transfer timesteps of 90 minutes \citep{leconte:2013a, kopparapu:2017} and 150 minutes \citep{way:2017} have been used by ExoCAM and other models successfully.  Even larger radiative transfer timesteps may remain valid for tidally locked planets, where the zenith angle does not change in time.  Still, anytime that model timesteps are increased in length, whether the primary, dynamical, or radiative, numerical instabilities become more common occurrences.  Lastly, a recent work by \citet{zhang:2021} used ExoCAM to develop an inverse modeling mode\footnote{https://github.com/Yixiao-Zhang/ evolving-CO2-in-ExoCAM} in 3D.  Inverse modeling, where the surface temperature is prescribed and the radiative forcings are allowed to evolve (greenhouse or solar), is commonly used in 1-D models \citep{kopparapu:2013}; however \citet{zhang:2021} applied the method to ExoCAM and was able to effectively duplicate the forward modeling results of \citep{wolf:2018a} for high-CO$_2$ atmospheres while reducing the computational burden per simulation by a factor of ten.

\section{Sensitivity to Moist Physical Parameters} \label{subsec:sensitivity}
A common criticism of 3D climate models used for exoplanet modeling is that they often contain tunable parameters that were originally calibrated to replicate the climate of the modern Earth and thus may be a source of uncertainty, leading to untrustworthiness of modeling results for extrasolar planets.  Of particular interest here, tunable parameters are common in the moist physics routines for treating sub-grid scale convection and clouds.  Horizontal grid resolutions for 3D climate models are generally large, with horizontal boxes spanning as much as several hundred kilometers on each side.  Thus unresolved small scale cloud and convection processes, such as localized cloud fractions, cloud overlap, and convective plumes, must be represented statistically through parameterizations.  In CESM, the values of tunable parameters that control these subgrid scale representations are dependent on the choices of horizontal resolution, dynamical core, and moist physics routines.  While the principle THAI simulations used the default model tunings for the selected setup (i.e. 4\degree$\times$5\degree grid, FV dynamical core, CAM4 moist physics), it is unknown if these tunings would necessarily be appropriate for an alien planet.  Thus an important part of building confidence in our 3D exoplanet climate models is not only in comparing them against other leading 3D models, but also in comparing them against themselves while exploring the effects of the many important assumptions and choices that are baked into out-of-the-box simulations.  {In this section we constrain how large an effect these tunable parameters may have on our subsequent climate predictions of TRAPPIST-1e.  While we acknowledge that this section does not constitute a comprehensive analysis of all temporal and spatial differences evident due to parameter sensitivities, our first order analysis supports the robustness of our 3D exoclimate modeling results against plausible changes in important tunable parameters.}  

We conduct a series of intramodel sensitivity tests with ExoCAM using the THAI protocol \citep{fauchez_thai:2019b} \textit{hab1} and \textit{hab2} cases as our starting points.  We refer the reader to the core series of THAI papers for more details on the THAI protocol and also for detailed analysis of the results \citep{fauchez_thai:2019b, turbet:2021, sergeev:2021, fauchez:2021b}.  To summarize the THAI protocols briefly here, both \textit{hab1} and \textit{hab2} cases simulate the climate of TRAPPIST-1e, having a radius of 0.91~R$_\earth$, a surface gravity of 0.93~R$_\earth$, a period of 6.1 Earth-days in synchronous rotation with a 2600~K host star, while receiving 900~Wm$^{-2}$ of stellar irradiation.  For both \textit{hab1} and \textit{hab2} cases, the planet is assumed to be completely ocean covered with an ocean albedo of 0.06 and a snow and sea ice albedo uniformly of 0.25.  The ocean is treated as a 100-meter deep slab ocean with no heat transport.  The atmospheric compositions used are a 1~bar N$_2$ dominated atmosphere with 400~ppm of CO$_2$ for \textit{hab1}, and a 1~bar CO$_2$ atmosphere for \textit{hab2}.  In each case water vapor, water clouds, and sea ice are allowed to evolve naturally in the system.   

\citet{sergeev:2021} discusses the two moist cases (\textit{hab1} and \textit{hab2}), contrasting the model results between ExoCAM, LMD-G, ROCKE-3D and UM for the standard THAI project intercomparison.  Here, including the two moist cases from the standard THAI project discussed in \citet{sergeev:2021}, we consider results from 43 ExoCAM simulations, iterating on \textit{hab1} and \textit{hab2}.  Table \ref{parametertable} summarizes the parameters for which sensitivity tests are conducted, including the default values for our model configuration and the ranges explored. The standard THAI simulations used the default values listed.  Here, we bracket these default values with end-member values drawn from CESM parameter selections used for other {model configurations that they offer.  Recall, ExoCAM is a patch for CESM that enables exoplanet configurations.  CESM independently contains numerous different model configurations spanning the pre-industrial through 21st century Earth, with different combinations of vertical and horizontal resolutions, model physics, and dynamical cores.  Each of the many configurations offered requires differing values for the subgrid scale tuning parameters in question.  Here we explore a range parameter values within the context of ExoCAM configuration, with plausible values informed by other existing CESM configuration requirements.}

Parameters explored include the cloud-overlap assumption, the liquid cloud droplet radius, cloud fraction relative humidity thresholds for low and high clouds, critical radii for autoconversion of liquid and ice condensate into precipitation, {the autoconversion coefficient for convective clouds, and the dynamical substepping frequency.  Because individual clouds cannot be resolved within large GCM grid boxes, their inherent horizontal patchiness and relation to adjacent cloud layers in the vertical dimension is parameterized through cloud fractions linked to cloud-overlap assumptions (e.g. \citet{pincus:2003}).  The cloud fraction represents the fraction of the horizontal grid-box that is covered by clouds, with the remaining portion assumed to be clear.  Cloud-overlap assumptions are used by the radiative transfer model to interpret the relative vertical layerings of clear and cloudy patches of sky.  Cloud fraction is generally controlled by simple scalings against the grid-box average relative humidity.  When the grid-box average relative humidity exceeds a certain threshold (typically $\sim$0.8 to 0.9) clouds begin to form with a cloud fraction that ramps towards unity as the grid-box average relative humidity approaches saturation.  Cloud droplet sizes and their subsequent conversion to precipitation are also treated via scaling considerations.  Autoconversion thresholds control the maximum size that typical cloud particles can grow before they become rain or snow, with separate thresholds for liquid, ice, and convectively produced clouds.  Autoconversion relies on the available amount of cloud condensate along with assumptions regarding the number density of cloud particles to determine when the size threshold to convert to precipitation is reached.  We also test fixing liquid cloud droplet sizes to different values which affects how fast they fall out of the atmosphere and strongly affects their radiative transfer properties.  Note that the default value of 14~$\mu$m in the \citet{rasch&kristjansson:1998} cloud scheme is taken from Earth observations of liquid water clouds over ocean waters, far from land.} 

Additionally, we conduct sensitivity tests using the native sea ice and snow albedos. Sea ice and snow albedos are divided into visible and near-infrared bands, and set to 0.67 and  0.3 for sea ice and 0.8 and 0.68 for sea ice and snow respectively in typical ExoCAM simulations following \citet{shields:2013}.  {Note that such two-band sea-ice and snow albedos would benefit from re-calibration according to the incoming stellar energy distribution of the specific star used (e.g. \citet{shields:2018}); nonetheless these default albedos provide a meaningful point of comparison against the spectrally uniform albedos specified in the THAI protocol \citep{fauchez_thai:2019b}.}  We also test the effects of using the native CAM5 cloud and convention physics (\citet{morrison&gettelman:2008, park&bretherton:2009, park:2014} hence forth MG clouds), instead of the CAM4 cloud and convection physics \citep{rasch&kristjansson:1998, hack:1994, zhang&mcfarlane:1995} which have been exclusively used in published ExoCAM works to date.  {As alluded to in Section \ref{subsec:generalfeatures}, the CAM5 cloud physics uses two-moment scheme that relies on time and spatial dependent aerosols to serve as sites for cloud condensation and thus can capture the microphysical interactions between clouds and aerosols with greater sophistication than does the CAM4 scheme.  Granted, this introduces another set of unknowns, as aerosols must be specified in context of an alien world.  Here, we use a default collection of time and spatial dependent aerosol fields derived for Earth to drive our CAM5 cloud microphysics.  We acknowledge that it is unlikely that TRAPPIST-1e would have an identical aerosol field as the Earth.}

Tables \ref{hab1table} and \ref{hab2table} present the global mean climatological results for \textit{hab1} and \textit{hab2} simulations respectively.  {Model outputs have been averaged over 10 Earth years ($\sim$600 TRAPPIST-1e orbits) after reaching thermal and radiative balance.}  As mentioned above, for the principle THAI project the \textit{n28archean} radiation scheme was used because development of the \textit{n68equiv} scheme was not completed in a timely manner.  {Henceforth these are referred to as the THAI standard cases.}  In this paper we focus on the results using our updated \textit{n68equiv} radiation scheme {as it provides more trustworthy results, particularly for the \textit{hab2} cases for reasons discussed in Section \ref{subsec:radiation}. Henceforth our nominal simululations with \textit{n68equiv} are referred to as our control cases.  Both THAI standard and control cases used default parameters selections from Table \ref{parametertable}.  Subsequent simulations listed in each of Tables \ref{hab1table} and \ref{hab2table} list the particular tuning parameter that was altered.}  Included in each table is the global mean surface temperature (T$_S$), the planetary albedo, and the integrated column amounts of water vapor, liquid water clouds, and ice water clouds.  All cases listed use the CAM4 cloud and convection physics except for the cases described as "MG clouds".  Note, that no sensitivity test was conducted using the native sea ice and snow albedos for \textit{hab2} because all sea ice and snow are trapped on the night-side of the planet, rendering any ice and snow albedo effect irrelevant due to the assumed synchronous rotation of TRAPPIST-1e.

\begin{figure}
  \centering
  \includegraphics[angle=0,width=8.3cm]{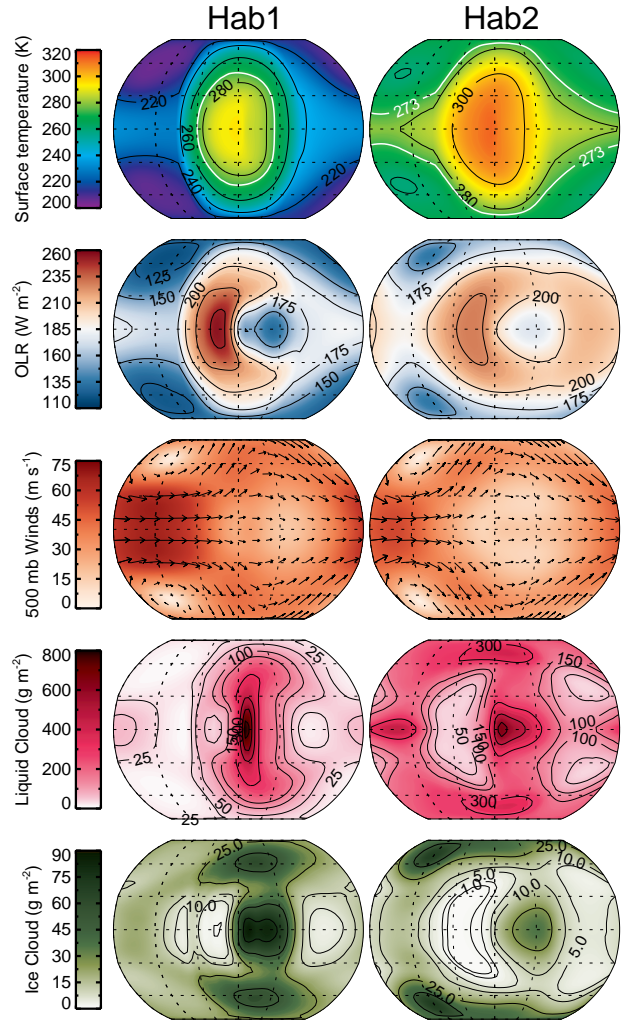}
  \caption{Contour plots showing surface temperature, outgoing longwave radiation (OLR), 500 mb vector winds, liquid cloud water, and ice cloud water for  \textit{hab1} (left) and \textit{hab2} simulations respectively.}
  \label{contour_fig}
\end{figure}

{Figure \ref{contour_fig} shows longitude-latitude contour plots for our \textit{hab1} and \textit{hab2} control simulations.  This figure provides a view of several important climatological variables; the surface temperature, outgoing longwave radiation (OLR), 500 mb vector winds, liquid water cloud column, and ice water cloud columns.  Note that the white line in the surface temperature plots outlines the approximate sea-ice margin.  Contour plots of these variables, and numerous other variables, for all \textit{hab1} and \textit{hab2} sensitivity simulations are included on an accompanying Zenodo repository\footnote{10.5281/zenodo.5532765}.  Output data of the time mean climates of all simulations are also provided on this repository.  
For both \textit{hab1} and \textit{hab2} the overall climate states and the dynamical regimes do not appear sensitive to our suite of tests.  A nominal range in global mean surface temperatures of $\sim$10~K are experienced in our sensitivity tests, with changing ice albedo and the liquid cloud droplet sizes generally providing the most significant feedbacks.  Still, the climate states of each set are not changed, with \textit{hab1} remaining cold and icy except around the substellar point, and \textit{hab1} remaining temperate.  A qualitative examination of the wind fields (see Zenodo supplement) of the \textit{hab2} simulations do not reveal any dramatic shifts in the dynamical regimes.  While the vector wind speeds can vary in magnitude, 200 and 500 mb vector wind fields appear qualitatively similar with vortices located on the night-side of the planet near the western terminator, strong supperrotation in the western hemisphere which then splits in two across to the eastern hemisphere.   A similar statement can be made for the \textit{hab1} cases, with the lone exception being the MG case.  This case displays the vortices in the eastern hemisphere, as opposed to the western hemisphere, and has weaker equatorial winds in the western hemisphere.  This change in dynamics is indicative of the importance of physics-dynamics couplings, and the role of clouds (and their radiative forcings) in influencing the general circulation and not just in being reactive to general circulation.  Note, that permitting the use of the Morrision-Gettlemen two-moment cloud scheme in ExoCAM is new feature which relies on the specification of a background aerosol field, that can depend on time and space.  I encourage others to explore this new model feature within ExoCAM more deeply.}

To better illustrate the resulting spreads in the climates of our sensitivity tests, we have collected global mean climatological information into Figures \ref{temperature_fig}, \ref{albemiss_fig}, \ref{watercloud_fig}, \ref{icecloud_fig}, \ref{cldvscld_fig}, \ref{cloudforcing_fig}, and \ref{greenhouse_fig}.  Each figure shows a scatter plot comparing two relevant climatological quantities along the x and y axes, with \textit{hab1} results in the left panel and \textit{hab2} results in the right panel.   {THAI standard simulations, featured in \citet{sergeev:2021}, are shown as a black "X" respectively; simulation number 1 in Tables \ref{hab1table} and \ref{hab2table}.  An orange cross indicates the revised THAI simulations using our updated radiative transfer scheme \textit{n68equiv}, which we take as our control simulation for further model sensitivity experiments in this work;} simulation number 2 in Tables \ref{hab1table} and \ref{hab2table}. Blue diamonds show the values for sensitivity tests {exploring different options and tuning parameters in the model.  Relative outlier simulations are labeled on each plot to add context; however generally our sensitivity tests cluster about the control case, implying that significant changes in the overall climate regimes cannot be driven by modifications to sub-grid scale tunable parameters.  Note the standard THAI simulations (e.g. \citet{sergeev:2021}) show insignificant deviation from our revised control cases for \textit{hab1}.  However, the standard THAI simulations show marked differences for \textit{hab2} cases in most plots.  This is a straight-forward consequence related to differences between the older radiative transfer version \textit{n28archean}, which overestimates both both near-infrared absorption by water-vapor under M-dwarf spectra and also the greenhouse effect from CO$_2$, as discussed in section \ref{subsec:radiation}.  The resulting climate for our THAI standard case is $\sim$11.5~K warmer than our control case, which feedbacks back on most climatologically interesting fields.  The results from using the \textit{n68equiv} yield a more trustworthy result, which is better in line with other models discussed in \citet{sergeev:2021}.  In interpreting these sensitivity simulations, the reader is reminded that tolerances for exoplanet climate science are considerably more forgiving than would be for Earth climate sciences.  For instance, several Kelvin differences in global mean temperature predictions, while enormous in the context of 21st century Earth, do not effect general conclusions regarding an exoplanet's climate state and potential for habitability.} 

\begin{figure}
  \centering
  \includegraphics[angle=0,width=14.5cm]{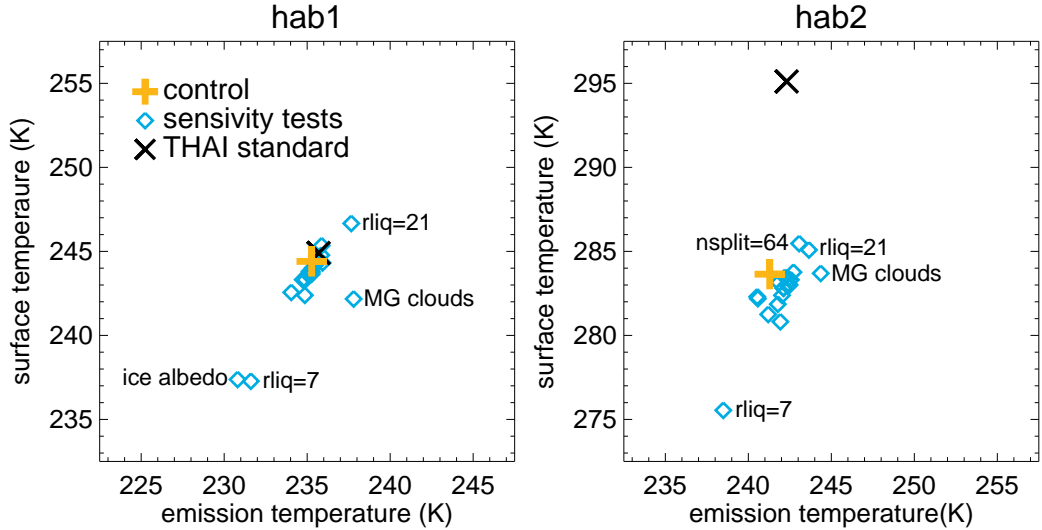}
  \caption{Sensitivity scatter plots for \textit{hab1} (left) and \textit{hab2} simulation sets.  Global mean surface temperature is plotted against the global mean emission temperature of the atmosphere.  The control simulation is indicated by the orange cross.  Sensitivity tests are indicated by blue diamonds with significant outliers labeled.}
  \label{temperature_fig}
\end{figure}

Figure \ref{temperature_fig} shows the global mean surface temperature (T$_s$) plotted versus the global mean emission temperature (T$_{emiss}$).  The emission temperature is calculated from a planet's global mean outgoing thermal radiation and inverting the Stefan-Boltzmann Law.  Comparing T$_s$ to T$_{emiss}$ yields information about the greenhouse effect of a planet.  Note that the x and y axes have different scales for left and right panels, but each axes spans 15~K.  Most parameter choices only have minor effects on both the T$_s$ and T$_{emiss}$.  There is significant clustering around the control for both \textit{hab1} and \textit{hab2} cases, featuring T$_s$ and T$_{emiss}$ differences of only $\sim$2~K.  Still, several significant outlier cases are noted in Figure \ref{temperature_fig}.  In particular the liquid cloud particle sizes (rliq), which are fixed to constant values in the radiation, have a meaningful impact on both the \textit{hab1} and \textit{hab2} climates.  Reducing the cloud particle size from 14~$\mu m$ to 7~$\mu m$ results in significant global mean cooling, up to $\sim$8~K, due to increased cloud scattering effects and decreased particle sedimentation, while increasing the cloud particle size to 21~$\mu m$ can increase the T$_s$ by several K.  Changes to cloud properties along with T$_s$ combine to affect the radiating level of the atmosphere, reflected by the values for T$_{emiss}$.  Naturally, the use of native ice albedos for \textit{hab1}, which are significantly higher than the 0.25 value used for the control case, cools the \textit{hab1} case significantly.  Interestingly, despite significant differences in methodology and in the water vapor and condensate partitioning (e.g. Figures \ref{watercloud_fig} and \ref{icecloud_fig}), the MG clouds cases have a relatively modest impact on T$_s$ and T$_{emiss}$.

\begin{figure}
  \centering
  \includegraphics[angle=0,width=14.5cm]{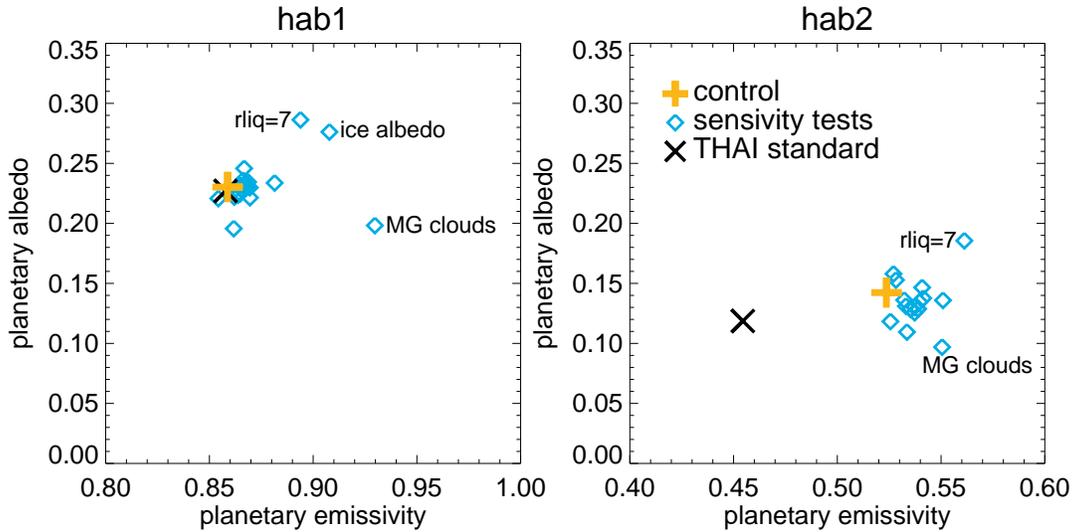}
  \caption{Sensitivity scatter plots for \textit{hab1} (left) and \textit{hab2} simulation sets.  Global mean column planetary albedo is plotted against the global mean planetary emissivity.  The control simulation is indicated by the orange cross.  Sensitivity tests are indicated by blue diamonds with significant outliers labeled.}
  \label{albemiss_fig}
\end{figure}

{In Figure \ref{albemiss_fig} we plot the global mean planetary albedo versus the planetary emissivity.  Here, we calculate the planetary emissivity as the ratio of the outgoing longwave radiative flux at the top-of-atmosphere to the outgoing radiative flux at the surface.  Using this simple formulation, a value of 1 indicates a planet with no greenhouse effect, while progressively lower planetary emissivities indicate a strengthening greenhouse effect.  The greenhouse effect is the totality of CO$_2$, water vapor, and cloud greenhouse effects.  Note that the y-axis (planetary albedo) are identical for both left and right panels, but the x-axis (planetary emissivity) have different scales.  Naturally, the \textit{hab2} simulations have a significantly lower emissivity than the \textit{hab1} cases due to their strong CO$_2$ greenhouse.  Notable outliers again are driven by the liquid cloud particle size, the MG cloud simulations, and the change in sea ice albedo.  While the mean surface temperature of the \textit{hab1} MG case does not deviate from the pack, Figure \ref{albemiss_fig} reveals that it has both a weaker greenhouse effect and a lower planetary albedo which offset each other.}

\begin{figure}
  \centering
  \includegraphics[angle=0,width=14.5cm]{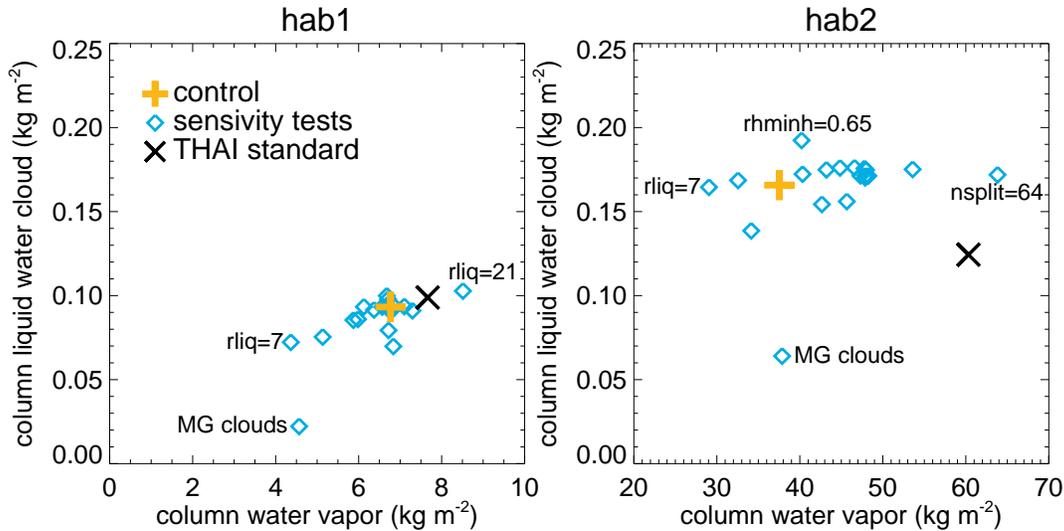}
  \caption{Sensitivity scatter plots for \textit{hab1} (left) and \textit{hab2} simulation sets.  Global mean column integrated liquid water cloud amount is plotted against the global mean column integrated water vapor amount.  The control simulation is indicated by the orange cross.  Sensitivity tests are indicated by blue diamonds with significant outliers labeled.}
  \label{watercloud_fig}
\end{figure}

\begin{figure}
  \centering
  \includegraphics[angle=0,width=14.5cm]{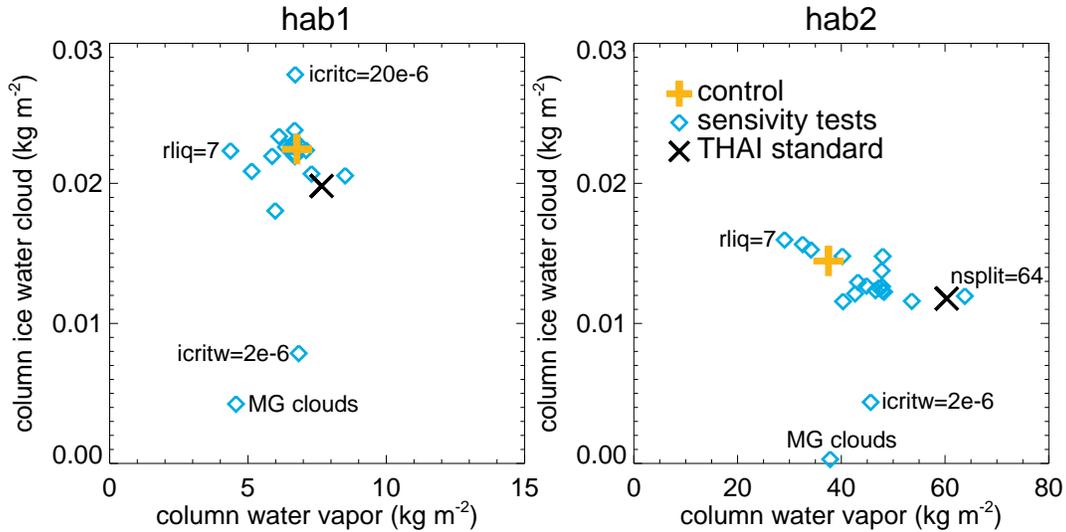}
  \caption{Sensitivity scatter plots for \textit{hab1} (left) and \textit{hab2} simulation sets.  Global mean column integrated ice water cloud amount is plotted against the global mean column integrated water vapor amount.  The control simulation is indicated by the orange cross.  Sensitivity tests are indicated by blue diamonds with significant outliers labeled.}
  \label{icecloud_fig}
\end{figure}

Figures \ref{watercloud_fig} and \ref{icecloud_fig} show global mean liquid and ice cloud column amounts plotted against the water vapor column amount, respectively.  In these figures note that cloud and water vapor columns are given in units of kg~m$^{-2}$, while in Tables \ref{hab1table} and \ref{hab2table}, cloud water columns are given in g~m$^{-2}$. Note that in Figures \ref{watercloud_fig} and \ref{icecloud_fig} the y-axis scales are identical, but the x-axis scales are different, because there is significantly less water vapor in the cold \textit{hab1} atmospheres compared to the temperate \textit{hab2} atmospheres that are warmed by 1~bar of CO$_2$.  By plotting the column cloud amounts against the column water vapor amounts, one gets a general idea of the partitioning of water in the atmosphere between thermodynamic phases.  In both Figures \ref{watercloud_fig} and \ref{icecloud_fig} for the \textit{hab1} cases we again see significant clustering around the control case.  Clustering is also evident for the \textit{hab2} sensitivity test simulations; however, the column water vapor amounts of the sensitivity tests are generally larger than is found in the control case.  The MG clouds cases are the largest outlier for both ice and liquid clouds amounts.  The MG cloud simulations have significantly less ice and liquid cloud water compared to the control in all cases, and generally have a water vapor column amount towards the lower end of the pack. Additionally, we find that various autoconversion size threshold parameters can have a meaningful effect of the cloud amounts.  Autoconversion tuning parameters control the particle size at which condensate is converted to precipitation and thus falls out of the atmosphere.  Smaller autoconversion thresholds increase precipitation rates and result in less cloud water remaining in the atmosphere.  Interestingly, note also that water vapor and clouds have a non-negligible sensitivity to the dynamical substepping frequency (\textit{nsplit}) particularly for the \textit{hab2} simulations.  {Thus, users must be cautious when ramping up the number of dynamical substeps to improve numerical stability mid-simulation, as it has a feedback on clouds and climate.}

{Figure \ref{cldvscld_fig} shows the global mean column integrated water amount, including both liquid and ice cloud total, comparing the substellar hemisphere mean value versus the antistellar hemisphere mean.  Interestingly, \textit{hab2} and \textit{hab1} simulations contain comparable amounts of substellar cloud water ($\sim$200~gm$^{-2}$).  However, the warmer \textit{hab2} climates experience more significant day-to-night transport of heat and water vapor, resulting in reduced day-to-night temperature gradients, and greatly increased atmospheric water vapor and cloud condensate on the night side of the planet.  These features can be clearly seen in Figure \ref{contour_fig}). }

\begin{figure}
  \centering
  \includegraphics[angle=0,width=14.5cm]{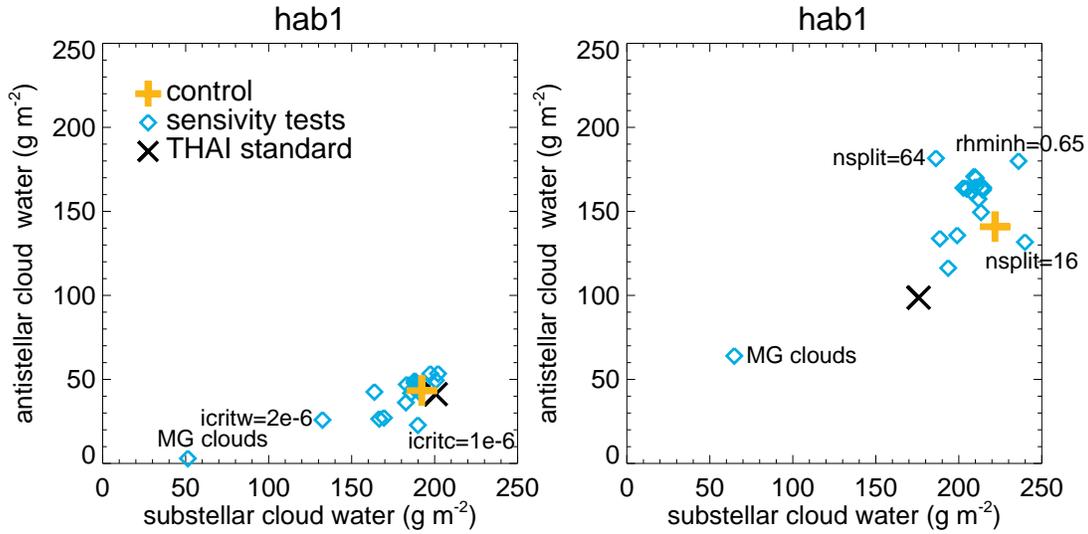}
  \caption{Sensitivity scatter plots for \textit{hab1} (left) and \textit{hab2} simulation sets.  Global mean column integrated water cloud amount (liquid + ice) is plotted for the substellar hemisphere mean value against the antistellar mean value.  The control simulation is indicated by the orange cross.  Sensitivity tests are indicated by blue diamonds with significant outliers labeled.}
  \label{cldvscld_fig}
\end{figure}

{We have also computed the global mean longwave and shortwave cloud radiative forcings produced in our runs.  In Figure \ref{cloudforcing_fig}, we plot the global mean shortwave cloud forcing versus the global mean longwave cloud forcing.  Shortwave cloud forcings are negative, as liquid water clouds on the day-side of the planet predominantly reflect stellar energy back to space, reducing the total amount of absorbed stellar radiation by the planet.  Longwave cloud forcings are positive, as high-alitutude ice clouds produce a greenhouse effect by lowering the effective emitting temperature of the planet.  Perhaps surprisingly, \textit{hab1} feature a larger magnitude shortwave forcing than does the \textit{hab2} cases.  One can observe in Figure \ref{contour_fig} that clouds are more concentrated right at the substellar point for \textit{hab1}, contributing to increased stellar reflection, whereas for the \textit{hab2} clouds are more evenly spread about the planet.  Both \textit{hab1} and \textit{hab2} clusters have similar longwave forcings. As expected, changing elements such as the liquid drop effective radii have noticeable effects on the cloud forcing fields. Similar to earlier cloud figures, the MG simulations stand as significant outliers compared to the rest with reduced magnitudes of both longwave and shortwave cloud forcings.  However, their reduction in magnitudes of cloud forcings approximately balances out, as their net cloud forcing falls among the main cluster as shown in Figure \ref{greenhouse_fig}.}

\begin{figure}
  \centering
  \includegraphics[angle=0,width=14.5cm]{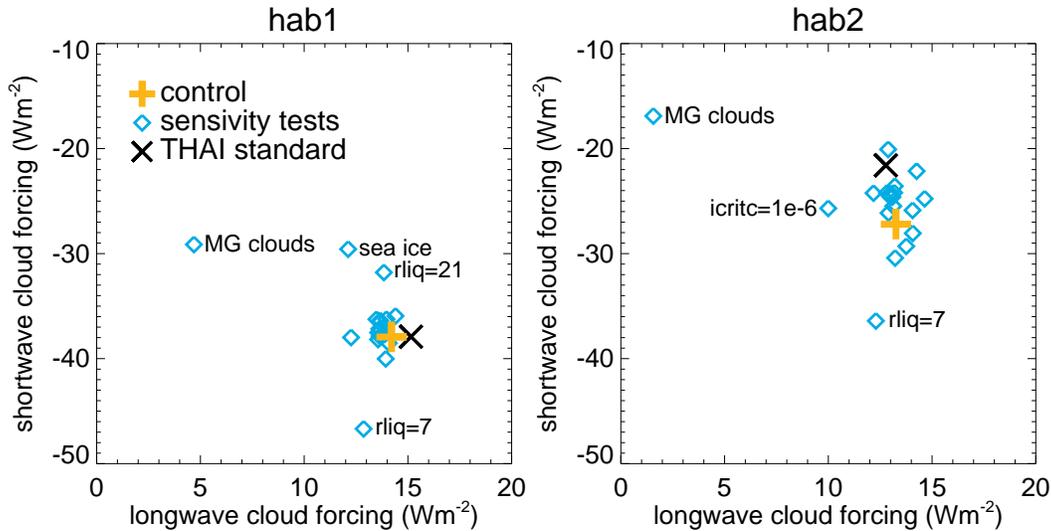}
  \caption{Sensitivity scatter plots for \textit{hab1} (left) and \textit{hab2} simulation sets.  The global mean shortwave cloud forcing is plotted against the global mean longwave cloud forcing.  The control simulation is indicated by the orange cross.  Sensitivity tests are indicated by blue diamonds with significant outliers labeled.}
  \label{cloudforcing_fig}
\end{figure}

{Lastly, Figure \ref{greenhouse_fig} plots the net cloud radiative forcings (shortwave + longwave) versus the clearsky greenhouse effect.  The clearsky greenhouse effect is computed as the difference between the upwelling longwave flux at the surface and the upwelling longwave flux at the top-of-atmosphere as calculated in parallel on-line clearsky radiative flux calculations.  Water vapor and CO$_2$ contribute to the clearsky greenhouse effect.  Naturally the magnitude of the clearsky greenhouse is much stronger for the \textit{hab2} cases, which have 1 bar CO$_2$ compared to the \textit{hab1} cases with only have 400 ppm of CO$_2$.  The water vapor greenhouse feedback also plays a role in amplifying the clearsky greenhouse effect in \textit{hab2} cases.  This water vapor amplification is perhaps best seen when comparing the \textit{hab2} control and THAI standard simulations, where the standard case has  Much of the additional warmth of the THAI standard case likely originates from the water vapor greenhouse feedback amplification of the original overestimation of the CO$_2$ greenhouse effect.}

\begin{figure}
  \centering
  \includegraphics[angle=0,width=14.5cm]{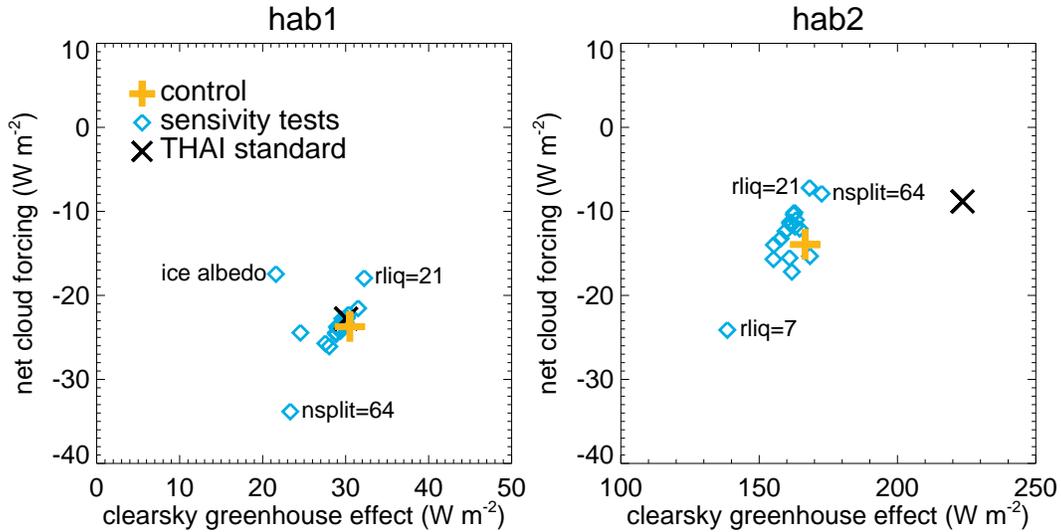}
  \caption{Sensitivity scatter plots for \textit{hab1} (left) and \textit{hab2} simulation sets.  The global mean shortwave net cloud forcing (longwave + shortwave) is plotted against the global mean clearsky greenhouse effect.  The control simulation is indicated by the orange cross.  Sensitivity tests are indicated by blue diamonds with significant outliers labeled.}
  \label{greenhouse_fig}
\end{figure}

In total, our results show that ExoCAM exhibits some sensitive to a variety of model parameter selections; however in the context of TRAPPIST-1e none of these selections, taken individually, dramatically change the climate outcomes and overall conclusions.  Unsurprisingly, bulk differences in the radiative transfer performance between \textit{n28archean} and \textit{n68equiv} can yield meaningful differences in climate results for {CO$_2$ dominated atmospheres.  Furthermore, a wholesale change in the model cloud schemes, from CAM4 to CAM5, produces significant deviations in cloud fields, but surprisingly little change to the mean surface temperature as changes to longwave and shortwave cloud forcings remarkably balance out.}  Among the free parameters studied here, notably assumptions for the surface albedos and {liquid cloud droplet particle sizes} play the largest roles in modulating climate.  {It is also possible that combinations of parameters could drive larger effects than are shown here taken individually.  However, such a multi-parameter study would be computationally expensive.}  Future 3D climate simulations could benefit from more refined treatments of surface albedos and improved cloud and aerosol microphysics.  


\section{Review of other sensitivity studies}

Note that other published works have also explored intramodel sensitivities for exoplanet modeling using ExoCAM and the NCAR family of models.  It is worth briefly reviewing these results here.  \citet{wei:2020} tested the climate sensitivity of ExoCAM under different horizontal and vertical resolutions for simulating tidally locked ocean covered planets around M-dwarf stars.  Their simulations show differences in T$_s$ of no more than $\sim$5~K for changes in horizontal resolution, and negligible differences in T$_s$ from changes in the vertical resolution.  Presumably, the CESM default resolution dependent tuning parameters were used, but these are not discussed in their work.  \citep{bin:2018} tested standard versions of CAM4 and CAM5 for calculating the inner edge of the habitable zone and compared their results to ExoCAM \citep{kopparapu:2017}.  \citet{bin:2018} found some notable differences between CAM4, CAM5, and ExoCAM results featured in \citep{kopparapu:2017}.  Note that the standard version of CAM5 uses the RRMTG radiative transfer package \citep{mlawer:1997} and the MG cloud scheme tested here, {although assumptions regarding aerosol fields are not discussed}.  In general, \citet{bin:2018} found that standard CAM5 performs similarly to ExoCAM in predicting the inner edge of the habitable zone for synchronously rotating M-dwarf planets.  However, the specific physical causes of these model deviations are not identified because there are simultaneous configuration differences in the radiation, clouds, convection, and boundary layer process between standard CAM5 and ExoCAM as used in\citet{kopparapu:2017}.  \citet{wolf&toon:2015} explored the model sensitivity of a legacy version of ExoCAM to cloud and convection tuning parameters for moist greenhouse climates possible for future Earth under high solar irradiation.  They found that liquid cloud particle sizes and cloud fraction relative humidity thresholds could drive changes of up to $\sim$10~K, while tuning of convective timescales generally had little effect.  \citet{yang_j:2013} conducted sensitivity tests for cool, slow rotating, tidally locked ocean covered planets around M-dwarf stars, for a variety of NCAR model configurations and parameters, similarly finding that the cloud liquid droplet sizes have the largest effect on climate.  Finally, \citet{rushby:2020} used ExoCAM to simulate TRAPPIST-1 planets as completely land covered planets, and found that the assumed surface albedo can have major ramifications for the global mean surface temperature of the planet. 

\section{Future Work and Goals}

ExoCAM has proven to be a useful community tool that has enabled numerous different research groups to study the climate and potential for habitability of extrasolar planets.  However, to ensure continued success, ExoCAM must continue to grow.  Presently, there are funded projects and collaborations to extend the functionality of ExoCAM to new worlds.  These will include expanding the current pressure-temperature spaces of the absorption coefficient tables along with adding new absorbing gas species in order to simulate a wider variety of planets, including oxygen-rich planets and sub-Neptune worlds.  Work is also currently ongoing to merge features from Mars-CAM \citep{urata&toon:2013a, urata&toon:2013b} with ExoCAM, including CO$_2$ condensation and the capability to model low pressure planets.  Other ongoing work will further develop the CARMA functionalities within ExoCAM, including fractal organic hazes \citep{wolf&toon:2010}, along with dust and ice clouds \citep{hartwick:2019}.  Finally, a collaborative effort will begin using ExoCAM with the dynamic ocean model (POP2) to study the long-term evolution of planetary climates in binary star systems.  All newly developed features will eventually be ported to the public ExoCAM and ExoRT versions on GitHub for community use, {as we believe strongly in the importance of provding useful tools for the community.  Where my energy and imagination wane, hopefully others can carry the science forward.}   

Of further note, one common user critique of ExoCAM is that it is not collected into a single location, but rather the user must download CESM separately (via SVN), and then download ExoCAM and ExoRT (via GitHub), and then go through several steps to combine the codes together.  However, starting with CESM version 2, CESM is now is available on GitHub and thus can be forked.  A long-term goal will be to create a forked version of CESM2 with ExoCAM and ExoRT fully integrated, so that users can obtain the entire code from a single location as a single code package.  However, such a large refactoring of the code would be time intensive, {and is unfortunately not on the near horizon.} 

\section{Summary}

ExoCAM is a branch of the National Center for Atmospheric Research Community Earth System Model version 1.2.1 that has been designed specifically for studying planetary and exoplanetary atmospheres.  Both CESM and ExoCAM are freely available to the community.  CESM is provided by NCAR, while ExoCAM is provided by the authors via GitHub as an easily applied model extension.  After its first public release in November of 2018, ExoCAM has been adopted by many groups in the community for simulating exoplanetary atmospheres.  As of the writing of this paper, ExoCAM has been used as a pillar in 37 peer reviewed papers, written by 20 different first authors, including numerous student led projects.  ExoCAM is also one of the core 4 climate models used in the TRAPPIST-1 Habitable Atmosphere Intercomparison project.  In this paper, we provided the first formal description of ExoCAM, along with details of a recent major upgrade to the radiative transfer module, ExoRT. 

In addition, here we have also tested ExoCAM for climate sensitivities to a variety of tunable parameters and model selections.  We have tested tunable parameters regarding clouds, convection, timestepping, and sea ice and snow albedos, along with model configurations regarding the radiative transfer scheme and moist physics packages.  Sensitivity tests were conducted in adherence with the THAI protocol \citep{fauchez_thai:2019b} for simulating habitable iterations of TRAPPIST-1e.  Changes to primary physical schemes, for instance the radiative transfer module or the cloud microphysics package, can yield meaningful changes to cloud amounts and the global mean surface temperature.  Changes to tunable parameters of the snow and sea ice albedos, or cloud particle sizes can also drive meaningful changes in the global mean surface temperatures.  However, the magnitude of these changes is never more than $\sim$10~K, indicating that basic climate states yielded for TRAPPIST-1e are reasonably robust for a variety of intramodel sensitivity tests.   Future sensitivity studies may want to explore if the combined sensitivities of multiple parameters changed simultaneously would yield larger perturbations to the climate.

In the coming years, there are several projects planned with ExoCAM that will expand its functionalities.  In particular, development will focus on expanding the radiative transfer capabilities to simulate a wider range of atmospheres, {adding CO$_2$ condensation}, and leveraging the CARMA cloud and aerosol microphysics package.  In continuing to push the model forward, we hope to continue to provide the community with a useful tool for studying planetary and exoplanetary atmospheres.

\acknowledgments

ETW, RK, and TJF acknowledge support from the GSFC Sellers Exoplanet Environments Collaboration (SEEC), which is supported by NASA’s Planetary Science Division’s Research Program. ETW, RK, and JHM acknowledge that this work was performed as part of NASA’s Virtual Planetary Laboratory, supported by the National Aeronautics and Space Administration through the NASA Astrobiology Institute under solicitation NNH12ZDA002C and Cooperative Agreement Number NNA13AA93A, and by the NASA Astrobiology Program under grant 80NSSC18K0829 as part of the Nexus for Exoplanet System Science (NExSS) research coordination network.  ETW, JHM, and RK acknowledge that this material is based upon work supported by the National Aeronautics and Space Administration (NASA) under Grant No. 80NSSC21K0905 issued through the Interdisciplinary Consortia for Astrobiology Research (ICAR) program.  JHM gratefully acknowledges funding from the NASA Habitable Worlds program under award 80NSSC20K0230.  ETW acknowledges that ExoCAM has been developed over many years and over the course of many grants that he has recieved. Thus, in addition to support from NASA's SEEC, VPL, and ICAR programs noted above, ETW also acknowledges the NASA Earth and Space Science Fellowship award NNX10AR97H, NASA Planetary Atmospheres Program award NNH13ZDA001N-PATM, and NASA Habitable Worlds awards NNX16AB61G, 80NSSC17K0741, 80NSSC17K0257, 80NSSC20K1421. All projects have provided critical funding for the ongoing development, maintenance, and support of ExoCAM and ExoRT.  ETW also acknowledges that this was his first, and perhaps last, lead author paper using Overleaf, and after this experience ETW has doubled-down on the thesis that Overleaf and \LaTeX\ are overhyped and overrated; however the co-authors know this thesis to be incorrect.



\bibliography{references}

%
%
\begin{table*}
\centering
\begin{tabular} {|c|c|c|c|c|c|c|c|}
\hline
\# & Short Name    & Description               & Default Value   & Sensitivity Tests \\
\hline
1 & cldovr         & cloud over lap scheme     & maximum-random & maximum, random \\
\hline
2 & rliq           & liquid cloud droplet radius ($\mu$m)      & 14    & 7, 21 \\
\hline
3 & rhminh         & cloud fraction relative humidity threshold, high      & 0.8     & 0.65, 0.90 \\
\hline
4 & rhminl         & cloud relative humidity threshold, low      & 0.9     & 0.85, 0.95 \\
\hline
5 & r3lcrit        & critical radius for autoconversion of liquid cloud drops (m)    & 10.0e-6     & 1.0e-6, 20.0e-6 \\
\hline
6 & zmconv$\_$c0   & autoconversion in coefficient, deep convection      & 0.0035     & 0.002, 0.045 \\
\hline
7 & icritc         & autoconversion of cold ice (m)       & 9.5e-6     & 1.0e-6, 20.0e-6 \\
\hline
8 & icritw         & autoconversion of warm ice (m)       & 2.0e-4     & 2.0e-6, 4.0e-4 \\
\hline
9 & nsplit         & dynamical substep frequency          & 32     & 16, 64 \\
\hline

\end{tabular}
\caption{Summary of tuning parameters used in the moist physics along with their default values and values used for sensitivity tests.}
\label{parametertable}
\end{table*}

%
%
\begin{table*}
\centering
\begin{tabular} {|c|c|c|c|c|c|c|c|}
\hline
\# & Radiation & Description                & T$_s$ (K) & albedo  & H$_2$O vapor (kg m$^{-2}$) & H$_2$O cld$_{liq}$ (g m$^{-2}$)  & H$_2$O cld$_{ice}$(g m$^{-2}$) \\
\hline
1 & n28archean & THAI standard              & 244.9  & 0.227  & 7.66  & 98.90  & 19.83 \\
\hline
2 & n68equiv   & control              & 244.4  & 0.230  & 6.77  & 93.24  & 22.44 \\
\hline
3 & n68equiv   & native sea ice albedo      & 237.3  & 0.276  & 5.13  & 75.41  & 20.87\\
\hline
4 & n68equiv   & cldovr = maximum           & 244.4  & 0.224  & 6.85  & 93.88  & 22.75 \\
\hline
5 & n68equiv   & cldovr = random            & 242.6  & 0.246  & 5.87  & 85.45  & 21.94 \\
\hline
6 & n68equiv   & icritc = 1.0e-6               & 242.4  & 0.234  & 5.98  & 85.93  & 18.04 \\
\hline
7 & n68equiv   & icritc = 20.0e-6              & 243.7  & 0.233  & 6.71  & 97.87  & 27.77 \\
\hline
8 & n68equiv   & icritw = 2.0e-6            & 245.3  & 0.221  & 6.83  & 69.81  & 7.87 \\
\hline
9 & n68equiv   & icritw = 4.0e-4            & 243.6  & 0.230  & 6.69  & 94.34  & 23.79  \\
\hline
10 & n68equiv  & MG clouds                  & 242.2  & 0.198  & 4.57  & 22.21  & 4.25  \\
\hline
11 & n68equiv  & r3lcrit = 1.0e-6              & 244.3  & 0.228  & 6.72  & 79.35  & 22.07  \\
\hline
12 & n68equiv  & r3lcrit = 20.0e-6             & 244.0  & 0.228  & 6.68  & 99.94  & 22.96 \\
\hline
13 & n68equiv  & rhminh = 0.65              & 243.3  & 0.236  & 6.12  & 93.34  & 23.35 \\
\hline
14 & n68equiv  & rhminh = 0.90              & 244.3  & 0.221  & 7.30  & 91.07  & 20.69  \\
\hline
15 & n68equiv  & rhminl = 0.85              & 243.3  & 0.234  & 6.37  & 91.53  & 22.59  \\
\hline
16 & n68equiv  & rhminl = 0.95              & 244.8  & 0.222  & 7.10  & 93.55  & 22.36 \\
\hline
17 & n68equiv  & rliq = 7                   & 237.4  & 0.286  & 4.37  & 72.27  & 22.33 \\
\hline
18 & n68equiv  & rliq = 21                  & 246.7  & 0.196  & 8.51  & 102.77 & 20.55 \\
\hline
19 & n68equiv  & zmconv$\_$c0 = 0.002       & 243.6  & 0.230  & 6.57  & 93.21  & 22.68 \\
\hline
20 & n68equiv  & zmconv$\_$c0 = 0.045       & 244.3  & 0.225  & 6.78  & 90.82  & 22.37 \\
\hline
21 & n68equiv  & nsplit = 16                & 243.8  & 0.230  & 6.70  & 92.59  & 21.93 \\
\hline
22 & n68equiv  & nsplit = 64                & 243.8  & 0.232  & 6.67  & 93.14  & 22.71 \\
\hline

\end{tabular}
\caption{Global mean climate statistics for \textit{hab1} simulations.  These simulations feature 1 bar N$_2$ plus 400 ppm CO$_2$ atmospheres for TRAPPIST-1e.}
\label{hab1table}
\end{table*}

%
%
\begin{table*}
\centering
\begin{tabular} {|c|c|c|c|c|c|c|c|}
\hline
\# & Radiation & Description                & T$_s$ (K) & albedo  & H$_2$O vapor (kg m$^{-2}$) & H$_2$O cld$_{liq}$(kg m$^{-2}$)  & H$_2$O cld$_{ice}$(kg m$^{-2}$) \\
\hline
1 & n28archean & THAI standard     & 295.1  & 0.119  & 60.32  & 124.39  & 11.78  \\
\hline
2 & n68equiv   & control     & 283.6  & 0.142  & 37.53  & 165.73  & 14.46 \\
\hline
3 & n68equiv   & cldovr = maximum  & 283.0  & 0.129  & 46.59  & 176.16  & 12.35 \\
\hline
4 & n68equiv   & cldovr = random   & 283.1  & 0.136  & 47.87  & 170.18  & 12.63 \\
\hline
5 & n68equiv   & icritc = 1.0e-6      & 280.8  & 0.136  & 40.33  & 172.31  & 11.57 \\
\hline
6 & n68equiv   & icritc = 20.0e-6     & 283.5  & 0.131  & 47.79  & 175.70  & 13.77 \\
\hline
7 & n68equiv   & icritw = 2.0e-6   & 282.9  & 0.129  & 45.66  & 156.06  & 4.38 \\
\hline
8 & n68equiv   & icritw = 4.0e-4   & 282.38 & 0.130  & 47.98  & 174.73  & 14.79 \\
\hline
9 & n68equiv  & MG clouds          & 283.7  & 0.097  & 37.86  & 64.05   & 0.30  \\
\hline
10 & n68equiv  & r3lcrit = 1.0e-6     & 282.2  & 0.153  & 34.14  & 138.60  & 15.26 \\
\hline
11 & n68equiv  & r3lcrit = 20.0e-6    & 282.8  & 0.131  & 47.30  & 171.41  & 12.56 \\
\hline
12 & n68equiv  & rhminh = 0.65     & 281.2  & 0.147  & 40.22  & 192.37  & 14.80 \\
\hline
13 & n68equiv  & rhminh = 0.90     & 283.8  & 0.129  & 42.65 &  154.35  & 12.12  \\
\hline
14 & n68equiv  & rhminl = 0.85     & 281.9  & 0.138  & 43.21  & 174.91  & 12.93 \\
\hline
15 & n68equiv  & rhminl = 0.95     & 283.3  & 0.126  & 48.26  & 171.28  & 12.25 \\
\hline
16 & n68equiv  & rliq = 7          & 275.5  & 0.186  & 29.05  & 164.54  & 15.97 \\
\hline
17 & n68equiv  & rliq = 21         & 285.1  & 0.109  & 53.59  & 175.17  & 11.59 \\
\hline
18 & n68equiv  & zmconv$\_$c0 = 0.002     & 282.4  & 0.134  & 44.85  & 176.01  & 12.68 \\
\hline
19 & n68equiv  & zmconv$\_$c0 = 0.045    & 283.4  & 0.129  & 47.91  & 171.31  & 12.32 \\
\hline
20 & n68equiv  & nsplit = 16              & 282.3  & 0.158  & 32.56  & 168.61  & 15.65 \\
\hline
21 & n68equiv  & nsplit = 64              & 285.5  & 0.118  & 63.82  & 171.90  & 11.95 \\
\hline

\end{tabular}
\caption{Global mean climate statistics for \textit{hab2} simulations.  These simulations feature 1 bar CO$_2$ atmospheres for TRAPPIST-1e.}
\label{hab2table}
\end{table*}

\end{document}